\input harvmac
\input epsf
\def\frak#1#2{{\textstyle{{#1}\over{#2}}}}
\def\frakk#1#2{{{#1}\over{#2}}}

\def\pa{\partial}
\def\semi{;\hfil\break}

\def\sy{supersymmetry}
\def\sic{supersymmetric}
\def\DRED{\ifmmode{{\rm DRED}} \else{{DRED}} \fi}
\def\DREDp{\ifmmode{{\rm DRED}'} \else{${\rm DRED}'$} \fi}
\def\NSVZ{\ifmmode{{\rm NSVZ}} \else{{NSVZ}} \fi}
  
\def\npb{{Nucl.\ Phys.\ }{\bf B}}   

\def\prl{Phys.\ Rev.\ Lett.\ }
\def\plb{{Phys.\ Lett.\ }{\bf B}}
\def\ijmpa{{Int.\ J.\ Mod.\ Phys.\ }{\bf A}}

\def\Ncal{{\cal N}}

\def\Wbar{\overline W{}}

\def\Phbar{\overline\Phi{}}

\def\Tr{\hbox{Tr}}

\def\hbar{{\overline h}{}}
\def \in{\leftskip = 40 pt\rightskip = 40pt}

\def \out{\leftskip = 0 pt\rightskip = 0pt}
{\nopagenumbers
\line{\hfil LTH 503} 
\line{\hfil hep-th/0105221}
\vskip .5in
\centerline{\titlefont Ultra-violet Finite Noncommutative Theories}
\vskip 1in
\centerline{\bf I.~Jack and D.R.T.~Jones}
\medskip
\centerline{\it Dept. of Mathematical Sciences,
University of Liverpool, Liverpool L69 3BX, UK}
\vskip .3in

We establish the  ultra-violet finiteness of various classes of 
noncommutative gauge theories.
  
\Date{May 2001}

There has been a great deal of recent interest in  noncommutative (NC) quantum
field theories,  stimulated by a connection with string theory and  
$M$-theory; see for example Refs.~\ref\filk{T.~Filk, \plb376 (1996) 53}%
\nref\Martruiz{
C.~P.~Martin and D.~Sanchez-Ruiz, \prl83 (1999) 476 [hep-th/9903077]}%
\nref\jab{M.M.~Sheikh-Jabbari, JHEP {\bf 9906} (1999) 015 [hep-th/9903107]}%
\nref\kraj{T.~Krajewski and R. Wulkenhaar, \ijmpa15 (2000) 1011
[hep-th/9903187]}%
\nref\bisu{
D.~Bigatti and L.~Susskind, Phys.\ Rev.\ D{\bf 62} (2000) 066004
[hep-th/9908056]
}%
\nref\haya{M.~Hayakawa, hep-th/9912167}%
\nref\MatusisJF{
A.~Matusis, L.~Susskind and N.~Toumbas,
JHEP {\bf 0012} (2000) 002
[hep-th/0002075]
}%
\nref\ZanonGY{
D.~Zanon,
Phys.\ Lett.\ B{\bf 504} (2001) 101
[hep-th/0009196]
}%
\nref\SantambrogioRS{
A.~Santambrogio and D.~Zanon,
JHEP {\bf 0101} (2001) 024
[hep-th/0010275]
}%
\nref\PerniciVA{
M.~Pernici, A.~Santambrogio and D.~Zanon,
Phys.\ Lett.\ B{\bf 504} (2001) 131
[hep-th/0011140]
}%
\nref\ZanonNQ{
D.~Zanon,
Phys.\ Lett.\ B{\bf 502} (2001) 265
[hep-th/0012009]
}%
\nref\FerraraMM{
S.~Ferrara and M.~A.~Lledo,
JHEP {\bf 0005} (2000) 008
[hep-th/0002084]
}%
\nref\ArmoniXR{
A.~Armoni,
Nucl.\ Phys.\ B{\bf 593} (2001) 229
[hep-th/0005208]
}%
\nref\ArmoniBR{
A.~Armoni, R.~Minasian and S.~Theisen,
hep-th/0102007
}%
\nref\TerashimaXQ{
S.~Terashima,
Phys.\ Lett.\ B{\bf 482} (2000) 276
[hep-th/0002119]
}%
\nref\RuizHU{
F.~R.~Ruiz,
Phys.\ Lett.\ B{\bf 502} (2001) 274
[hep-th/0012171]
}%
\nref\MartinDQ{
C.~P.~Martin and D.~Sanchez-Ruiz,
Nucl.\ Phys.\ B{\bf 598} (2001) 348
[hep-th/0012024]
}%
\nref\BonoraGA{
L.~Bonora and M.~Salizzoni,
Phys.\ Lett.\ B{\bf 504} (2001) 80
[hep-th/0011088]
}%
--\ref\BonoraTD{
L.~Bonora, M.~Schnabl, M.~M.~Sheikh-Jabbari and A.~Tomasiello,
Nucl.\ Phys.\ B{\bf 589} (2000) 461
[hep-th/0006091]
}. The theories have, moreover, novel properties which make
them  worthy of attention in their own right;  for example NC quantum
electrodynamics  exhibits both asymptotic  freedom and charge
quantisation. 

The algebra of functions  on a noncommutative space 
is isomorphic to the algebra of functions on a commutative 
space with coordinates $x^{\mu}$, with the product $f*g (x)$ 
defined as follows
\eqn\prodnc{
f*g (x) = e^{-i\Theta^{\mu\nu}
\frakk{\pa}{\pa\xi^{\mu}}\frakk{\pa}{\pa\eta^{\nu}}}
f(x+\xi)g(x+\eta)|_{\xi,\eta\to 0},}
where $\Theta$ is a real antisymmetric matrix.
Quantum field theories analogous to the corresponding commuting 
theories are  now straightforward to define, with $*$-products replacing 
ordinary products. In the case of 
gauge theories there are a number of subtleties, however. 
Consider a field $\phi(x)$ which transforms as follows under 
a local symmetry transformation:
\eqn\untrans{
\phi(x) \to \phi'(x) = U(x)*\phi(x) = e^{i\Lambda(x)}_* *\phi (x),}
where 
\eqn\suncase{
e^{i\Lambda(x)}_* = 1 + i\Lambda - \frakk{1}{2!}\Lambda*\Lambda+\cdots}
By considering the product $U_1*U_2 = e^{i\Lambda_1}_* * e^{i\Lambda_2}_*$
it is easy to show that $SU_N$ is not a group under the $*$-product, 
whereas $U_N$ is, so that we will devote our 
attention to $U_N$ gauge theories. Such gauge theories are constructed 
using the gauge fields $A_{\mu}$ and matter fields $\chi, \xi, \phi$ 
(scalars or fermions)
transforming as follows:
\eqna\redefg$$\eqalignno{
A'_{\mu} &= U*A_{\mu}*U^{-1} + ig^{-1}U*\pa_{\mu}U^{-1}&\redefg a\cr
\chi'  &= U*\chi &\redefg b\cr
\xi' &= \xi*U^{-1} &\redefg c\cr
\phi'  &= U*\phi*U^{-1} &\redefg d\cr}$$
where $\chi, \xi, \phi$ transform according to the fundamental, 
the anti-fundamental and the adjoint representations respectively. 
One may also, of course, have matter singlets; but, as has been noted 
by previous authors, it is not clear how to construct 
other representations (such as fractionally charged particles in the 
$U_1$ case).

In this paper we consider the ultra-violet (UV) divergences of NC theories,
and in particular seek theories that are UV finite.
Consider the pure (no matter) $U_N$ NC gauge theory (NCGT). 
If one computes the
one loop corrections and isolates the UV divergence, one
finds that this can be described both for $N=1$\Martruiz\jab\kraj\haya\
and for $N\geq 2$ \jab\kraj\ArmoniXR\BonoraGA\ by a single $\beta$-function
$\beta_g$, which is moreover identical  (for $N\geq 2$) 
to the corresponding one-loop $\beta_g$
for the $SU_N$ commutative theory (CGT). 
(Contrast this to the  $U_N$ CGT case, where 
of course, writing $U_N\equiv SU_N\otimes U_1$, 
the $U_1$ gauge coupling is unrenormalised). 
Although our chief interest here is in supersymmetric 
theories, an elementary consequence of our methods 
is that for the pure $U_N$ gauge theory, the NCGT $\beta_g$ is  
{\it to all orders\/} identical to the large $N$ approximation to the 
corresponding $SU_N$ CGT $\beta_g$.   

The NC formalism extends readily to \sic\ theories\foot{Note in 
particular that in Ref.~\PerniciVA\ the gauge invariance of the one loop
effective action for the ${\cal N =} 4$ theory was demonstrated}. 
An ${\cal N} =1$ $U_N$ gauge theory 
with a set of adjoint chiral superfields $\Phi_i$ is described 
by the Lagrangian 
\eqn\nsusy{
L = \int d^4\theta\, \Tr\left( e_*^{-gV}*\Phbar_i *e_*^{gV}*\Phi_i\right)
+\left[\int d^2\theta
\left(W(\Phi_i)+\frak14 W^{\alpha}*W_{\alpha}\right)+\hbox{c.c.}\right],}
where $V$ is the vector superfield, $W^{\alpha}$
the corresponding field strength, and the superpotential $W(\Phi_i)$ 
is holomorphic and gauge invariant.

We will focus particularly on the following two theories:
\eqn\wone{
W_1 = h_1\Tr\left( \Phi_1*\left[\Phi_2,\Phi_3\right]_*\right)=h_1(W_a-W_b)}
\eqn\wtwo{
W_2 = h_2\Tr\left( \Phi_1*\left\{\Phi_2,\Phi_3\right\}_*\right)=h_2(W_a+W_b)}
where $W_a=\Tr(\Phi_1*\Phi_2*\Phi_3)$ and $W_b=\Tr(\Phi_1*\Phi_3*\Phi_2)$, 
and $\Phi_{1\cdots3}$ are adjoint chiral supermultiplets. If we define 
\eqn\vecref{
\Phi = \frakk{1}{\sqrt{2}}\phi^a\lambda^a, \quad a=0,1,\cdots N^2 -1}
where $[\lambda^a, \lambda^b] = 2if^{abc}\lambda^c,
\{\lambda^a, \lambda^b\} = 2d^{abc}\lambda^c$, 
and $\Tr(\lambda^a\lambda^b) = 2\delta^{ab}$,
then in the commutative versions of the above theories we would have
\eqn\wonec{W^C_1 = i\sqrt{2}h_1f^{abc}\phi_1^a\phi_2^b\phi_3^c}
and  
\eqn\wtwoc{W^C_2 = \sqrt{2}h_2d^{abc}\phi_1^a\phi_2^b\phi_3^c}
and it is interesting to contrast this with the NC case where we 
have 
\eqn\wonenc{W_1 = \frakk{h_1}{\sqrt{2}}
\left( d^{abc}\phi_1^a*\left[\phi_2^b,\phi_3^c\right]_* + 
if^{abc}\phi_1^a*\left\{\phi_2^b,\phi_3^c\right\}_*\right)}
and  
\eqn\wtwonc{W_2 = \frakk{h_2}{\sqrt{2}}
\left( d^{abc}\phi_1^a*\left\{\phi_2^b,\phi_3^c\right\}_* +
if^{abc}\phi_1^a*\left[\phi_2^b,\phi_3^c\right]_*\right)}

In both the CGT and the NCGT cases, $W_1$ corresponds to ${\cal N} = 4$ 
\sy, if we set $h_1 = g$.  It is well-known that the ${\cal N} = 4$ CGT  
is all orders finite\foot{The ${\cal N}=4$ $U_N$ CGT 
consists of the direct product of the familiar ${\cal N}=4$ $SU_N$  
theory with a ${\cal N}=4$ $U_1$ free field theory}; 
as we shall see the same is true 
in the NCGT ${\cal N} = 4$ case. This is to be expected since in 
general NC theories have improved UV divergence properties.
Somewhat more surprising, however, is the following:
in the CGT case, the $SU_N$ version of $W_2^C$, 
for the case 
\eqn\jmmodel{h_2 = gN/\sqrt{N^2-4}} is the so-called 
${\cal N} = 4d$ model discussed in 
Refs.~\ref\jome{D.R.T. Jones and L. Mezincescu, \plb 138 (1984) 293},
\ref\pawe{A.J. Parkes and P.C. West, \npb 256 (1985) 340}. 
It is UV finite through   two loops, but has a three (and higher) 
loop divergence 
\ref\JonesAY{
D.R.T.~Jones and A.J.~Parkes,
Phys.\ Lett.\ B{\bf 160} (1985) 267
}, which can, however, be removed
\ref\JonesVP{
D.R.T.~Jones,
Nucl.\ Phys.\ B{\bf 277} (1986) 153\semi
A.V.~Ermushev, D.I.~Kazakov and O.V.~Tarasov,
Nucl.\ Phys.\ B{\bf 281} (1987) 72\semi
D.I.~Kazakov,
Phys.\ Lett.\ B{\bf 179} (1986) 352
}
by replacing Eq.~\jmmodel\ by
\eqn\jmmodelb{h_2 = gN/\sqrt{N^2-4)} + a_5 g^5 + \cdots}
where $a_5,\cdots$ are calculable constants.   
In the NCGT case the $U_N$ version of the theory is, as we shall see, 
all orders UV finite
simply given $h_2 = g$, in other words  
without recourse to the kind of coupling constant redefinition represented 
by Eq.~\jmmodelb. 

Since the $\Phi$ are adjoint fields in $U_N$ we can use the diagrammatic 
notation 
originally introduced by 't Hooft
\ref\tHooftJZ{
G.~'t Hooft,
Nucl.\ Phys.\ B{\bf 72} (1974) 461\semi
P.~Cvitanovic, P.G.~Lauwers and P.N.~Scharbach,
Nucl.\ Phys.\ B{\bf 203} (1982) 385
}, where we represent $\Phi^a_b$
by a double line as in Fig.~1, the arrow pointing towards the upper index. 
This is in fact a considerable simplification compared to the 
generalised $f^{abc}, d^{abc}$ formulation that has been used in some 
papers. 
\vskip 20pt
\epsfysize= 0.25in
\centerline{\epsfbox{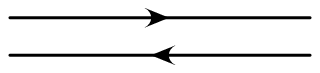}}
\vskip 10pt\in
{\it \noindent Fig.~1:
The propagator for an adjoint $U(N)$ field}
\bigskip
\out
The vertices $W_a$, $W_b$, and their complex conjugates
$\overline{W}_a$ and  $\overline{W}_b$
are then represented as in Fig.~2. 
\bigskip
\epsfysize= 2.5in
\centerline{\epsfbox{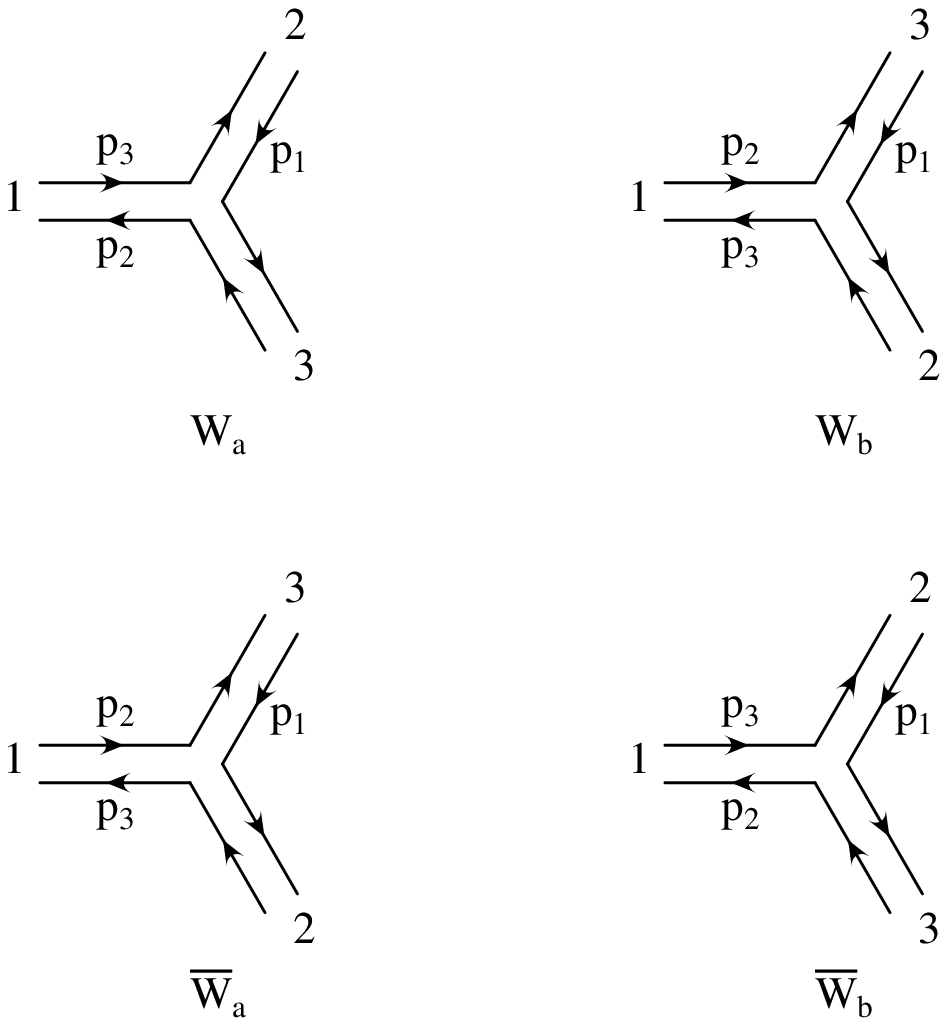}}
\in
{\it \noindent Fig.~2:
The vertices $W_a$, $W_b$, $\overline{W}_a$, $\overline{W}_b$.}
\bigskip
\out
In momentum space,
$W_a$ is associated with a factor $e^{ik_1\wedge k_2}$ where $k_i$ is the
momentum associated with $\Phi_i$ and $p\wedge q=\Theta^{\mu\nu}p_{\mu}
q_{\nu}$. Suppose we associate momenta $p_i$ with the lines as
shown in Fig.~2 (flowing in the direction of the arrows), 
so that for $W_a$, $k_1=p_3-p_2$ etc, and for $W_b$,
$k_1=p_2-p_3$ etc. Then the exponential factor for $W_a$ can be rewritten 
using \eqn\exfac{
k_1\wedge k_2=p_1\wedge p_2+p_2\wedge p_3+p_3\wedge p_1}
as 
\eqn\exfaca{
e^{i\sum_{\rm legs}p_{\rm out}\wedge p_{\rm in}}=\prod_{\rm legs}
e^{ip_{\rm out}\wedge p_{\rm in}}}
where $p_{\rm out}$, $p_{\rm in}$ are~the momenta associated with the lines 
with arrows pointing out from, or into, the vertex respectively for each leg. 
We thereby 
associate an exponential factor with each leg of the vertex. It is easy to 
check that the exponential factor can also be written in the form Eq.~\exfaca\
for $W_b$ and indeed for 
$\Wbar_a=\Tr\left(\Phbar_1*\Phbar_3*\Phbar_2\right)$ and 
$\Wbar_b=\Tr\left(\Phbar_1*\Phbar_2*\Phbar_3\right)$. 
Moreover the $\Phbar_i\Phi_iV^n$ vertex is given by the expression
$\Tr\left({1\over{n!}}
\Phi_i[[\ldots[[\Phbar_i,V]_*,V]_*\ldots V]_*,V]_*\right)$ with $n$ nested
commutators. Again, the exponential factor for one of these vertices can be 
written in the form given in Eq.~\exfaca. 

We claim that it is only planar graphs 
constructed using the vertices above which contribute to the 
renormalisation-group (RG) functions 
($\beta$-functions and anomalous dimensions)
for the theories with $W_1$ or $W_2$ in the noncommutative case. 
Let us start by considering the theory with $W_1$. 
Consider for example the one loop contribution
to the anomalous dimension of $\Phi_1$ given by contracting $W_1$ with 
$\Wbar_1$. The contractions of $W_a$ with $\Wbar_a$, or $W_b$ with
$\Wbar_b$, give planar diagrams, as depicted in Figs~3(a,b), 
while the contractions of $W_a$ with 
$\Wbar_b$, or $W_b$ with $\Wbar_a$, give non-planar graphs, as depicted in
Fig.~3(c,d). 
\bigskip
\epsfysize= 2.0in
\centerline{\epsfbox{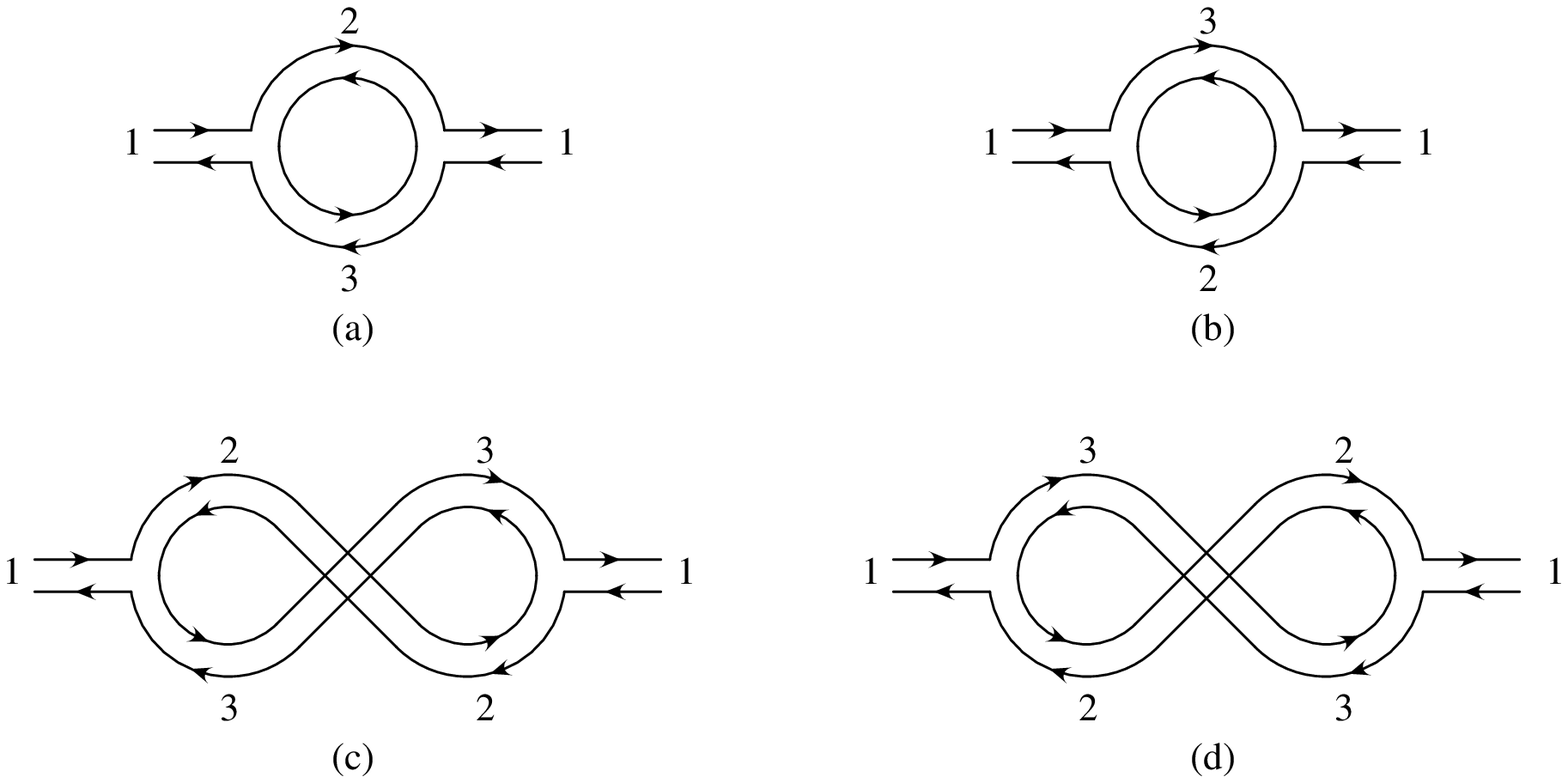}}
\in
{\it \noindent Fig.~3:
The one-loop diagrams.}
\bigskip
\out
Now these four diagrams all correspond to the same
one-loop momentum integral with a single loop momentum. For the planar graphs
in Figs.~3(a,b), the loop momentum may be assigned to the closed loop and 
momenta may be assigned to the other lines consistently with momentum
conservation at the two vertices and a given external momentum. 
It is then clear from Eq.~\exfaca\ that the exponential factors on the internal 
pairs of lines cancel in pairs; because the ``out'' momentum for one vertex is 
the ``in'' momentum for its neighbour. The remaining exponential factors from
the external legs cancel by momentum conservation. In particular there is no 
phase factor containing the loop momentum which, if present, would
suppress the ultraviolet divergence\filk--\bisu. On the 
other hand, in the case of the non-planar graphs in Figs.~3(c,d), there is no 
closed loop to which the loop momentum can be assigned, 
the above argument breaks down, and therefore there will be a phase
factor involving the loop momentum (as can easily be checked) making the 
diagram ultra-violet finite.

This argument readily extends to higher loop orders, to graphs
containing gauge fields, and to other RG-functions. For any planar graph, the 
loop momenta from the
corresponding Feynman graph may be assigned to closed loops of the
planar graph, and the  exponential factors cancel in pairs on internal
pairs of lines. In the case of the non-planar graphs, there are fewer
planar loops (of the kind apparent in Figs.~3(a,b))
than loop momenta and this argument breaks down. There will then
be an overall exponential factor involving at least one of the loop
momenta, and this graph will not contribute to the RG-function. Of
course a non-planar graph (with a phase factor)  may have a planar (and
hence divergent) sub-graph, but this graph will be finite  after
subtraction of sub-divergences; this is analogous to the way that in
commutative $\phi^4$ theory, the $\phi^6$ 1PI Green's function  is
finite, in spite of the fact that it includes 4-point sub-graphs. 

We now turn to the theory with superpotential $W_2$. We shall show that its
divergences are the same as those of the theory with superpotential $W_1$
(with $h_1\rightarrow h_2$). The difference between the superpotentials
$W_1$ and $W_2$ (apart from $h_1\rightarrow h_2$) 
lies simply in the sign of $W_b$ (and $\Wbar_b$). 
For simplicity we start with diagrams which 
only contain Yukawa vertices. Note that of course by chirality
$W$s and $\overline{W}$s must alternate in such a diagram. 
Once again the only divergent diagrams are the planar ones. Consider any
planar diagram. We may assign it an odd or even ``parity'' according as its sign
is changed or unchanged by $W_b\rightarrow-W_b$, $\Wbar_b\rightarrow-\Wbar_b$.
We would like to show that every planar diagram has even parity. For 
simplicity, 
suppose we join together the external legs of the diagram and imagine it to be
drawn on the surface of a sphere. 
We see that for planar diagrams every closed loop has the same sense of 
rotation for the arrow (anti-clockwise with our conventions). Therefore the 
fields 
$\Phi_1$, $\Phi_2$ and $\Phi_3$ always appear in clockwise order for $W_a$ 
and anticlockwise order for $W_b$; and conversely, the fields $\Phbar_1$, 
$\Phbar_2$ and $\Phbar_3$ always appear in anticlockwise order for $\Wbar_a$ 
and clockwise order for $\Wbar_b$. 
\bigskip
\epsfysize= 1.0in
\centerline{\epsfbox{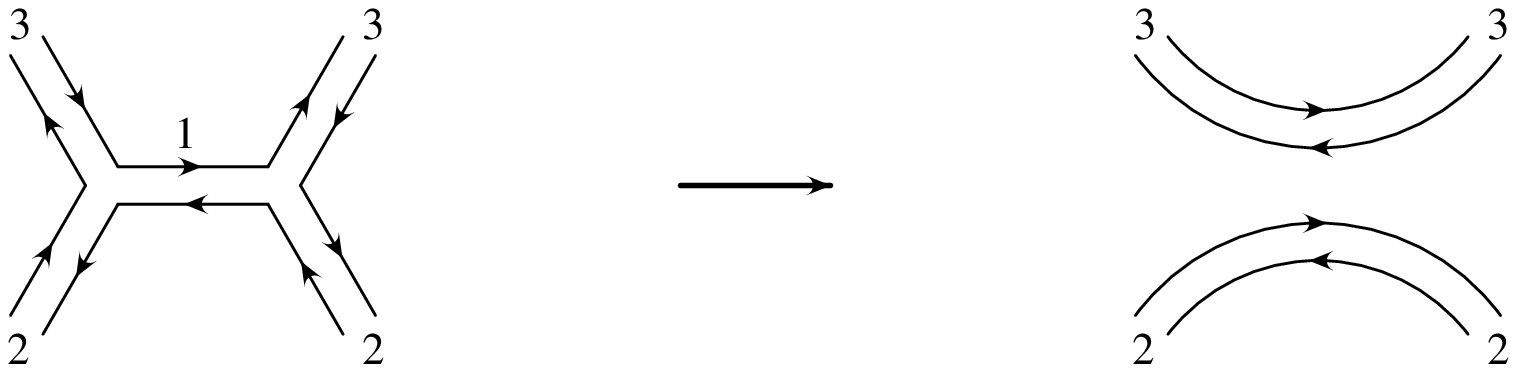}}
\in
{\it \noindent Fig.~4:
Reduction of diagrams.}
\bigskip
\out
If the diagram contains a pair of linked adjacent vertices $W_a$ and $\Wbar_a$,
or $W_b$ and $\Wbar_b$, (signalled by a sequence of fields such as 
$\Phi_2\Phi_1\Phi_2$ in some loop) 
then we may obtain a graph with two fewer vertices 
and the same parity by the process depicted in Fig.~4. 

We now repeat this 
process until we can do so no further. The 
process could terminate in one of two ways:
the first possibility is that eventually we obtain a diagram consisting of 
separate closed 
loops and no vertices, which clearly has even parity by default; and thus 
the original diagram must have even parity. The second  
possibility is that eventually every loop 
consists of a permutation of the sequence $\Phi_1\Phi_2\Phi_3$ repeated
an integral number $n$ times (where $n$ would be even by chirality). But it is
easy to see that this is impossible for planar diagrams. The diagram would then 
consist entirely of
hexagons, dodecagons and so on. Suppose we have a diagram with $n_6$ 
hexagons, $n_{12}$ dodecagons etc. Let $n_V$ be the number of vertices,
$n_P$ the number of propagators and $n_L$ the number of loops. Then we have
\eqn\eula{\eqalign{
3n_V=&6n_6+12n_{12}+18n_{18}+\ldots\Rightarrow
 n_V=2n_6+4n_{12}+6n_{18}+\ldots,\cr
2n_P=&6n_6+12n_{12}+18n_{18}+\ldots\Rightarrow 
n_V=3n_6+6n_{12}+9n_{18}+\ldots,\cr
n_L=&n_6+n_{12}+n_{18}+\ldots,\cr}}
and then
\eqn\eulb{
n_V-n_P+n_L=-n_{12}-2n_{18}-\ldots}
so that Euler's formula 
\eqn\eulg{n_V-n_P+n_L=2-2{\cal G}} 
has no solution for the 
sphere which has genus ${\cal G} = 0$. 
We deduce that the second possibility does not in fact occur, 
and therefore the original 
diagram is indeed of even parity. It is easy to extend this argument to
graphs with gauge propagators, by noting that we may remove a gauge
propagator without changing the parity of the graph. 
It follows that the divergences, and thus
the RG-functions, of the theory with superpotential $W_2$ may be obtained from 
those for superpotential $W_1$ by replacing $h_1$ with $h_2$. 

Our main results now follow immediately upon setting $h_1=h_2=g$. Firstly, 
the theory with $W_1$ now becomes ${\cal N}=4$ NCGT. 
So we see that the $\Ncal=4$ NCGT $\beta$ functions are derived from the planar
graphs. We now note that these planar graphs are exactly those which
give the leading  $N$ contribution to the $\beta$-function for the
$\Ncal=4$ CGT, since at  $L$ loops they contain the maximum number ($L$)
of closed loops. Since the $\Ncal=4$ CGT is finite, the leading $N$
contributions must vanish individually  at each loop order. Therefore
the $\Ncal=4$ NCGT $\beta$ functions must also vanish, and 
$\Ncal=4$ NCGT is finite to all orders. Secondly, since the RG-functions for 
the theory with $W_2$ are identical to those of the theory with $W_1$, the
theory with $W_2$ is also finite to all orders (for $h_2=g$).  

Clearly the fact that both $W_1$ and $W_2$ lead (for $h_1 = h_2 = g$) to 
finite theories, and the obvious similarity between Eq.~\wonenc\ 
and Eq.~\wtwonc, 
suggest that $W_2$ also represents a theory with ${\cal N} > 1$ \sy;
however we have been unable to demonstrate this. The presence of 
the commutator in Eq.~\wone\ (as opposed to the anti-commutator in 
Eq.~\wtwo) is crucial for the additional symmetries 
(as given, for example, in 
Ref~\ref\GrisaruNK{
M.~Grisaru, M.~Rocek and W.~Siegel,
Phys.\ Rev.\ Lett.\  {\bf 45} (1980) 1063
}) associated 
with the ${\cal N} = 4$ invariance. It would clearly be 
interesting to compare the two theories in the infra-red; 
it has been argued\MatusisJF\ that the ${\cal N} =4$ theory is free 
of divergences as $\Theta\to 0$, although such divergences 
are characteristic of NC theories in general. 

By similar reasoning we can use the finiteness of commutative 
${\cal N} = 2$ theories 
beyond one loop to establish the corresponding result in the NCGT case, 
that is for superpotentials of the form 
\eqn\neqtwo{
W = h\sum_{n=1}^{N_f} \xi_n * \Phi * \chi_n}
where $\xi_n, \phi, \chi_n$ transform according to the superfield 
generalisation of Eq.~\redefg{}, 
and for ${\cal N} = 2$ \sy\ we require $h = \sqrt{2}g$. 
The contributions to RG-functions are associated once again 
with cancellation of phase factors in planar graphs; here these 
contributions are (at $L\geq 2$ loops) precisely given by the terms 
of order $N^L, N^{L-1}N_f, N^{L-1}N_f^2, \cdots N N_f^{L-1}$ from 
the corresponding RG-functions for the CGT.  (This corresponds to the 
Veneziano 
\ref\VenezianoWM{
G.~Veneziano,
Nucl.\ Phys.\ B{\bf 117} (1976) 519
}\
(as opposed to the 't Hooft) limit, i.e. both  $N, N_f\to\infty$ with 
$N/N_f$ fixed.)  
Since in the CGT case 
the RG-functions vanish beyond loop, it follows that all these contributions 
cancel. By choosing 
$N_f = 2N$ for one loop finiteness we obtain another class of 
all orders UV finite theories. 

In conclusion: we have established the UV finiteness to all orders 
of the ${\cal N} = 4$ $U_N$ NCGT, a closely related ${\cal N} = 1$
theory and the class of one-loop finite ${\cal N} = 2$  $U_N$ 
theories. A simple corollary of our 
methods is that $\beta_g$ for the pure non-\sic\ $U_N$ NCGT is identical to 
the large-$N$ (or planar) approximation to the $\beta_g$ for 
the corresponding $SU_N$ CGT; and for NCQCD (with $N_f$ flavours) 
the $L$-loop contribution to 
the $U_N$ RG-functions are given by the 
terms from the corresponding commutative $SU_N$ QCD
RG-functions of the form $N^a N_f^b$ where $a + b = L$, corresponding once 
again to the Veneziano\VenezianoWM\ limit.

\vfill
\eject
\line{\it Note Added\hfil}
After this paper was submitted to the archive we were made aware of some
related work: 

The UV/IR connection (the existence of infra-red singularities
arising from large virtual momenta) was 
described in Ref.~\ref\MinwallaPX{
S.~Minwalla, M.~Van Raamsdonk and N.~Seiberg,
JHEP {\bf 0002}  (2000) 020
[hep-th/9912072]
}. This paper deals mainly with scalar theories, and in fact
describes the cancellation of phase factors involving internal 
momenta in planar 
graphs by use of momentum assignments like those shown in our Fig.~2. 
A rigorous proof of renormalisability for various 
massive  NC scalar theories (in particular $\phi^*\phi\phi^*\phi$ 
for $d=4$) was given in Ref.~\ref\ChepelevHM{
I.~Chepelev and R.~Roiban,
JHEP {\bf 0103} (2001) 001
[hep-th/0008090]
}.
The relevance of the Veneziano limit  
for NCQCD described  above was remarked in Ref.~\ref\AmbjornNB{
J.~Ambjorn, Y.~M.~Makeenko, J.~Nishimura and R.~J.~Szabo,
Phys.\ Lett.\ B{\bf 480} (2000) 399 
[hep-th/0002158]
}.
A general proof of the renormalisability of a particular 
supersymmetric noncommutative theory is given for the Wess-Zumino 
model in 
Ref.~\ref\riv{H.O.~Girotti, M.~Gomes, V.O.~Rivelles and
A.J.~da Silva, \npb587 (2000) 299
[hep-th/0005272]
}.
It was pointed out in Ref.~\bisu\ and re-emphasised in 
Ref.~\ref\arm1{A.~Armoni, JHEP {\bf 0003} (2000) 033
[hep-th/9910031]
} 
that the divergences
of pure $U_N$ noncommutative gauge theory are dictated by the large $N$ 
limit of the commutative theory. The latter paper also raises the interesting
possibility of finite, possibly non-supersymmetric  noncommutative theories 
obtained by orbifold truncation of the ${\cal N}=4$ theory. 
We also
mention the possibility of defining finite noncommutative theories on 
fuzzy spheres\ref\stein{H.~Grosse, 
C.~Klimcik and P.~Presnajder, Comm.~Math.~Phys.~180 (1996) 429
[hep-th/9602115]
; Int.~J.~Theor.~Phys.~35 (1996) 231
[hep-th/9505175]
}.
(See Ref.~\ref\steina{H.~Grosse, J.~Madore and H.~Steinacker, 
J.~Geom.~Phys.~38 (2001) 308 
[hep-th/0103164]
}
for the $q$-deformed case.)

\bigskip\centerline{{\bf Acknowledgements}}\

We thank Hugh Morton for his crucial assistance and patient explanation of
matters graphical. One of us (IJ) thanks the Laboratory of Mathematics and 
Theoretical Physics at the Fran\c cois Rabelais University in
Tours for hospitality while part of this work was carried out.

\listrefs
\bye